\documentclass{elsart}

\usepackage{amssymb}
\usepackage{graphicx}

\begin{document}
\begin{frontmatter}



\title{Search for strange tribaryon states in the inclusive $^4$He($K^-_{stopped},~p$) reaction}


\author[titech]{M.~Sato\thanksref{RIKEN}\ead{m-sato@riken.jp}},
\author[snu]{H.~Bhang},
\author[tsu]{J.~Chiba},
\author[snu]{Seonho~Choi},
\author[titech]{Y.~Fukuda},
\author[tsu]{T.~Hanaki},
\author[ut]{R.~S.~Hayano},
\author[riken]{M.~Iio},
\author[ut]{T.~Ishikawa},
\author[kek]{S.~Ishimoto},
\author[smi]{T.~Ishiwatari},
\author[riken]{K.~Itahashi},
\author[kek]{M.~Iwai},
\author[titech,riken]{M.~Iwasaki},
\author[smi,tum]{P.~Kienle},
\author[snu]{J.~H.~Kim\thanksref{kriss}},
\author[riken]{Y.~Matsuda},
\author[riken]{H.~Ohnishi},
\author[riken]{S.~Okada},
\author[riken]{H.~Outa},
\author[kek]{S.~Suzuki},
\author[riken]{T.~Suzuki},
\author[riken]{D.~Tomono},
\author[smi]{E.~Widmann},
\author[ut,riken]{T.~Yamazaki},
\author[snu]{H.~Yim}

\address[titech]{Department of Physics, Tokyo Institute of Technology, Tokyo 152-8551, Japan}
\address[snu]{Department of Physics, Seoul National University, Seoul 151-742, South Korea} 
\address[tsu]{Department of Physics, Tokyo University of Science, Chiba 278-8510, Japan}
\address[ut]{Department of Physics, The University of Tokyo, Tokyo 113-0033, Japan}
\address[riken]{RIKEN Nishina Center,RIKEN, Saitama 351-0198, Japan}
\address[kek]{High Energy Accelerator Research Organization (KEK), Ibaraki 305-0801, Japan}
\address[smi]{Stefan Meyer Institut f\"{u}r subatomare Physik, \"{O}sterreichische Akademie der Wissenschaften,  A-1090 Wien, Austria}
\address[tum]{Physik Department, Technische Universit\"{a}t M\"{u}nchen, D-85748 Garching, Germany}
\thanks[RIKEN]{Present address: RIKEN Nishina Center, RIKEN, Saitama 351-0198, Japan}
\thanks[kriss]{Present address: Korea Research Institute of Standards and Science (KRISS), Daejeon 305-600, South Korea} 

\begin{abstract}
A search for tribaryon states was performed at KEK-PS. 
We adopted the $^4$He($K^-_{stopped},~p)$ reaction to populate the states with strangeness $-$1, charge 0 and isospin 1. 
No significant narrow structure was observed in the mass region from 3000 to 3200 MeV/$c^2$ in an inclusive missing mass spectrum. 
The upper limit of the formation branching ratio was determined to be 
 ($1\sim3) \times 10^{-4}$, ($0.7\sim2)\times 10^{-3}$ and ($2\sim8) \times 10^{-3}$/($stopped~K^{-}$) with 95 \% confidence level for narrow states with an assumed width of 0, 20 and 40 MeV/$c^2$, respectively.

\end{abstract}

\end{frontmatter}

\section{Introduction}
A possible existence of a tribaryon state with strangeness $-1$ and isospin 0 has been predicted theoretically by Akaishi and Yamazaki as a deeply-bound $\overline{K}$ nuclear state in $^3$He nuclei \cite{AY02}.  
Recently, the KEK-PS E471 group reported a distinct peak
 in a missing mass spectrum of the $^4$He($K^{-}_{stopped},~p)$ reaction \cite{E471p}, which was interpreted to be a signal of a tribaryon state with strangeness $-$1 and isospin 1, called S$^0(3115)$ (M = 3117.0$^{+1.5}_{-4.4}$ MeV/$c^2$ and $\Gamma~ <$ 21 MeV/$c^2$). 
 The same group obtained another candidate of the tribaryon state, S$^{+}$(3140), in the $^4$He($K^{-}_{stopped},~n)$ missing mass spectrum, but 
its statistical significance was not sufficient to claim a definitive evidence for its existence \cite{E471n}.
These results  have triggered many theoretical interpretations \cite{ADY05,Nona,OT}. 
Meanwhile the FINUDA group observed a peak in the proton momentum spectrum of $^6$Li($K^-_{stopped},~p$) \cite{Finuda}. 
They attributed the peak to the proton emission from the $K^-$ capture on a ``quasi'' deuteron in $^6$Li as, $K^{-}~d~\rightarrow p~{\it \Sigma^{-}}$ ($P_{p}=488$ MeV/$c$), which was introduced as an interpretation of S$^0$(3115) by Ref. \cite{OT}, and they claimed  that the peak structures observed in E471 and FINUDA should be  of the same nature. 
Counter arguments to this interpretation followed \cite{AY07}.
Therefore, the confirmation and further study of these states were strongly awaited.

This situation has lead us to carry out a new experimental search with improved  resolution and higher statistics based on the E471 experimental setup.
The reaction we adopted is $^4$He($K^{-}_{stopped},~N$), the same as in E471;
\begin{eqnarray*}
K^-_{stopped} + \mathrm{^4He} \rightarrow \mathrm{S} + N,
\end{eqnarray*}
where the state with isospin 1 and charge 0 (S$^0$) is populated in the proton emission reaction and the state with isospin 1 and 0 and charge $+$1 (S$^+$) state in the neutron emission reaction.
The state originally predicted by Akaishi and Yamazaki has be searched for using neutron spectroscopy.

The objectives of the proton spectroscopy are, first, to confirm S$^0(3115)$ and determine its width and formation ratio with improved resolution by an inclusive measurement. Since the experimental setup of E471 was originally designed to be optimized for the neutron TOF measurement, the experimental resolution for protons was not satisfactory.
Moreover, the limited momentum acceptance for protons made the search for excited states of S$^{0}(3115)$ difficult.
Thus, the second objective of the present experiment is to search for other candidates for tribaryon states, including excited states of S$^0(3115)$.

In this Letter, we give the first report, namely, on experimental results on the missing mass analysis of the $^4$He($K^{-}_{stopped},~p$) reaction, and results on the neutron measurement will be given in a paper to follow.

\section{Experimental Method}
\begin{figure}[!bht]
\begin{center}
\includegraphics[width=\columnwidth]{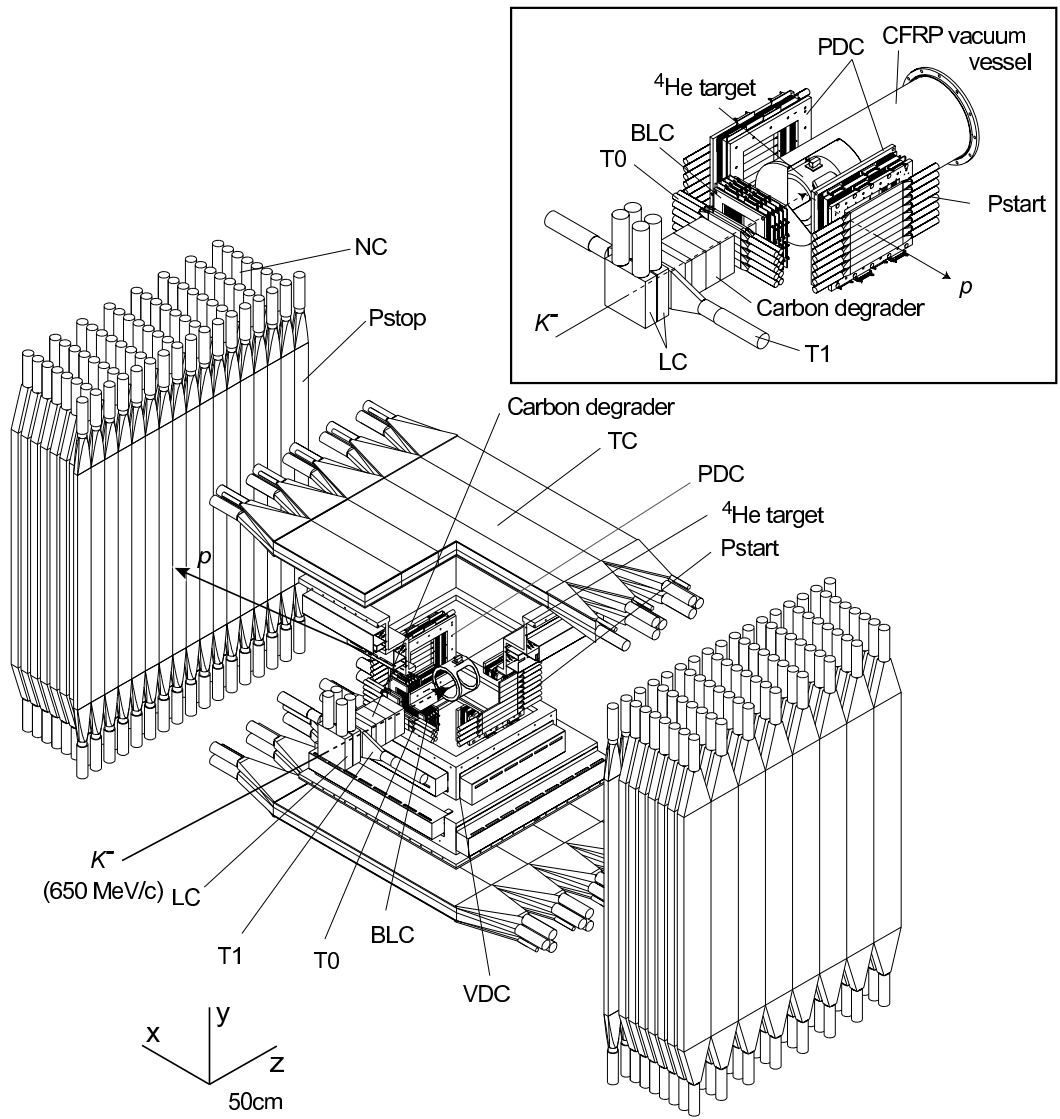}
\caption{The partially cutaway view of the E549 experimental setup. The inset figure shows the closeup of the beamline apparatus.}
\label{setup}
\end{center}
\end{figure}
The experiment (called ``E549'') was performed at the K5 beamline of the KEK 12 GeV proton synchrotron. The experimental setup is schematically illustrated in Fig. \ref{setup} together with a detailed view around the target in the inset.
A $K^-$ beam with momentum 650 MeV/$c$ was stopped inside a super-fluid helium-4 target (20 cm in diameter, 15 cm long at a density of 0.145 g/cm$^3$). 
We determined the energy of protons emitted from the $K^-$ absorption reaction by means of the time-of-flight (TOF) method.

The experimental apparatus was based on that of the former experiment E471, and its description can be found in Ref \cite{E471p,E471nim}.
Since the setup of E471 was optimized for neutron spectroscopy, considerable improvements have been made to achieve better energy resolution for the proton TOF measurement. 

In E549, TOF detectors dedicated for the proton measurement were installed. They consist of a tracking drift chamber (PDC) and a set of TOF start/stop counters ($P_{start}$ and $P_{stop}$), which are plastic scintillator walls with 8 and 17 segments, respectively. 
The missing mass resolution was expected to be twice that of E471 thanks to the precise determination of the TOF and the flight path.  Beside this, the $^4$He($K^{-}_{stopped},~p$) data were taken inclusively, whereas the proton data in E471 were measured in coincidence with secondary charged particles (``semi-inclusive''). The present inclusive measurement enables us to determine the formation ratio of the state precisely.

Together with the beam timing counter (T0), $P_{start}$ also gives us information on the reaction timing which is defined  by the time difference between the stopped  kaon  and outgoing charged particles. 

Furthermore, we extended the momentum acceptance for the proton to perform the experimental search with a wider missing mass region than in E471. This became possible by minimizing the amount of material between the target and $P_{stop}$.
 
Data were taken for a month from the end of May, 2005, and the accumulated data correspond to (1.03 $\pm$ 0.15) $\times~10^8$ stopped $K^-$ inside the target.

\section{Analysis and Results}
\subsection{Determination of the TOF resolution}
There is a well-known fact that a few percent of stopped $K^-$'s in liquid helium form  long-lived atomic states with  large angular momenta. 
These ``{\it metastable}''  kaonic He atoms were directly observed in a time spectrum by a past KEK experiment with $K^-$ absorption on $^4$He \cite{meta}. 
Since the partial lifetime of this state ($\tau_{meta}$=59$\pm$4 ns\footnote{The lifetime and fraction values reported in Ref. \cite{meta} were revised in Ref. \cite{OutaD} after refined analysis.}\cite{OutaD}) is much longer than the kaon lifetime, most kaons in this state decay at-rest without nuclear absorption.  
A monochromatic muon associated with a two body decay of stopped $K^-$ ($K^{-}\rightarrow\mu^{-}\overline{\nu}_{\mu}$, called $K^{-}_{\mu2}$) was used for determining the timing resolution of the present TOF system and its long-term stability throughout the experiment. 
The yield of $K^-_{\mu2}$ was also used for the normalization of the missing mass spectrum as described later. 

\begin{figure}[hbt]
\begin{center}
\includegraphics[width=0.5\columnwidth]{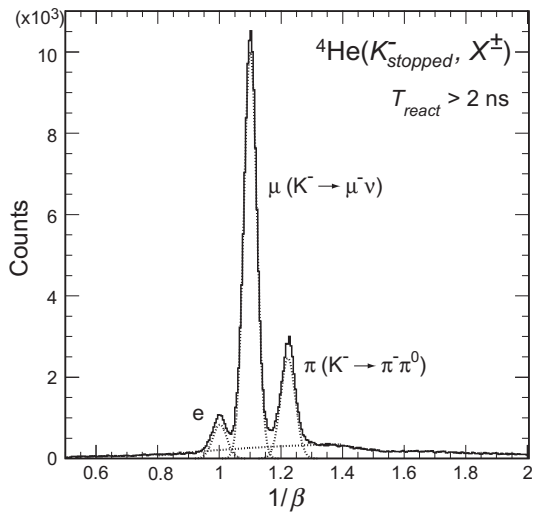}
\caption{1/$\beta$ spectrum from the $^4$He($K^{-}_{stopped}$, $X^{\pm}$) reaction with the delayed event selection. The spectrum was fitted by a three Gaussian and a second-order polynomial function. Fitting results are overlaid by dotted lines.}
\label{meta}
\end{center}
\end{figure}

These decay components, especially $K^{-}_{\mu2}$, can  be clearly identified by applying the timing analysis between the stopped kaons and the outgoing muons.  
This is because the lifetime of a kaon in the metastable atomic state together with the kaon decay lifetime ($\tau_{K^-}=10.24\pm0.11$ ns \cite{OutaD}) is much longer than the typical hyperon lifetimes which are of the order of 100 ps.
Thus, by selecting the delayed timing events, we can drastically suppress the events from the prompt nuclear absorption and successive hyperon decays.
Figure \ref{meta} shows the inverse velocity (1/$\beta$) spectrum of secondary charged particles from the $^4$He($K^{-}_{stopped},~X^{\pm}$) reaction with delayed timing selection ($T_{react}>2$ ns).
We can clearly see three peaks at $1/\beta = 1.0, 1.1, 1.24$. They correspond to  electrons (mainly $K^-_{e3}$), muons ($K^-_{\mu2}$) and pions ($K^-_{\pi2}$), respectively.
We estimated the TOF resolution from the width of the $K^{-}_{\mu2}$ muons, and found it to be 0.020 ($\sigma$) after fitting the spectrum with three Gaussians and a second-order polynomial function. This value is equivalent to that obtained from  $K^{+}_{\mu2}$ events in the stopped $K^+$ calibration data.

\subsection{Inclusive proton momentum spectrum}
\begin{figure}[bh]
\begin{center}
\includegraphics[width=0.5\columnwidth]{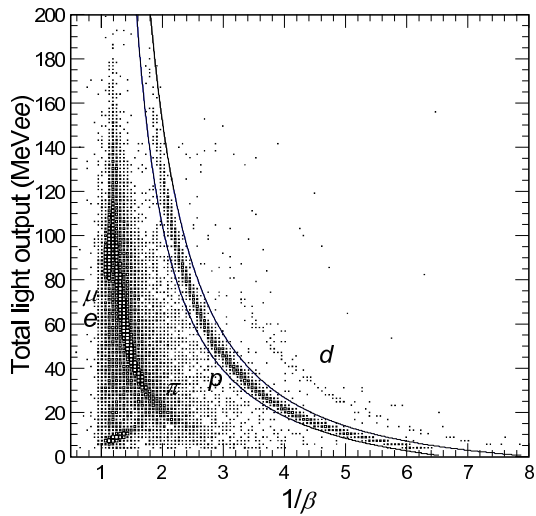}
\caption{Plot of the correlation of 1/$\beta$ and the total light output in $P_{stop}$ and NC in units of MeV$ee$ (MeV electron equivalent). The two lines in the figure define the proton region.}
\label{pid}
\end{center}
\end{figure}

Protons from the $^4$He($K^{-}_{stopped},~X^{\pm}$) reaction were clearly identified by the correlation between 1/$\beta$ and the total light output in P$_{\mathrm{stop}}$ and NC. 
 Figure \ref{pid} shows the criteria for proton selection. This selection method is essentially the same as that of E471, but  we can detect slower protons, i.e. a higher missing mass region,  thanks to the extension of the momentum acceptance (see Fig. 2 in Ref. \cite{E471p}).
The contamination due to reacted pions was estimated to be about 5 \% of protons in the present proton selection.

\begin{figure}[hbt]
\begin{center}
\includegraphics[width=0.5\columnwidth]{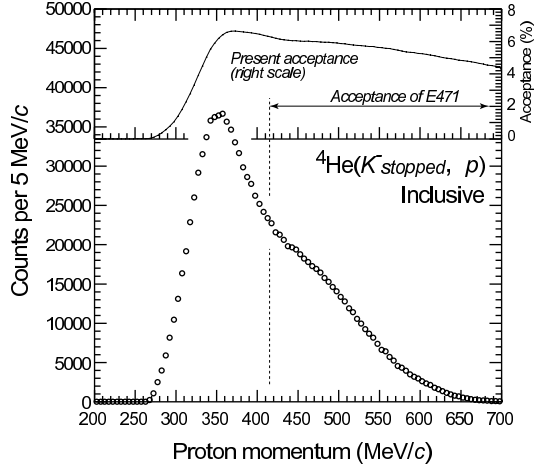}
\caption{Inclusive proton momentum spectrum from the $^4$He$(K^{-}_{stopped},~p)$ reaction. The upper figure  represents the geometrical acceptance for protons as a function of the proton momentum. The vertical line in the figure shows the acceptance limit of E471.}
\label{mom}
\end{center}
\end{figure}

Figure \ref{mom} shows the obtained momentum distribution under the inclusive condition.
In the figure, a correction for the energy loss between the reaction vertex and the TOF counters was applied. 
The overall spectrum shape is understood well by the two nucleon absorption reaction, $KNN\rightarrow YN$, for the higher momentum side and quasi-free hyperon production and its decay, $KN\rightarrow Y\pi$ and $Y\rightarrow N\pi$, for the lower one.
The upper figure shows the geometrical acceptance of protons, which 
 was obtained by a Monte Carlo simulation using the GEANT 4 package taking into account the realistic stopped $K^-$ distribution and the E549 setup.
The vertical line in the figure shows the lower acceptance limit in E471.
The present sensitive region was extended below 415 MeV/$c$. 
As shown in the figure, there is no distinct peak. In the following sections, we will determine the upper limit of the formation branching ratio for a tribaryon state.

\subsection{Normalization and missing mass spectrum}
The proton counts in the missing mass spectrum were normalized by the number of stopped $K^-$.
For the normalization, we used the metastable state described above, because the fraction of free decays, $f_{decay}$, of stopped $K^-$ on $^4$He  was measured to be 3.5 $\pm$ 0.5 \%/($stopped~K^{-}$) in Ref. \cite{OutaD}.
The proton yield, $Y_p$, is given as 
\begin{eqnarray*}
Y_{p}= \frac{ f_{decay} \cdot B_{K^{-}_{\mu2}}\cdot R \cdot\epsilon_{\mu} }{\epsilon_{p} \cdot N_{\mu}}  \cdot  N_{p}, 
\end{eqnarray*}
where $N_{\mu}$ and $N_p$ are the number of counts of the $K^{-}_{\mu2}$ muons and protons respectively,  
$B_{K^{-}_{\mu2}}$ is the branching ratio of the $K^{-}_{\mu2}$ decay, 
$R$ is the reduction factor resulting from the delayed timing gate, i.e. $\exp(-T_{react}/\tau_{K^-})$, and $\epsilon_{p}$ and $\epsilon_{\mu}$ are detection efficiencies of muons and protons respectively. 

\begin{figure}[hbt]
\begin{center}
\includegraphics[width=0.5\columnwidth]{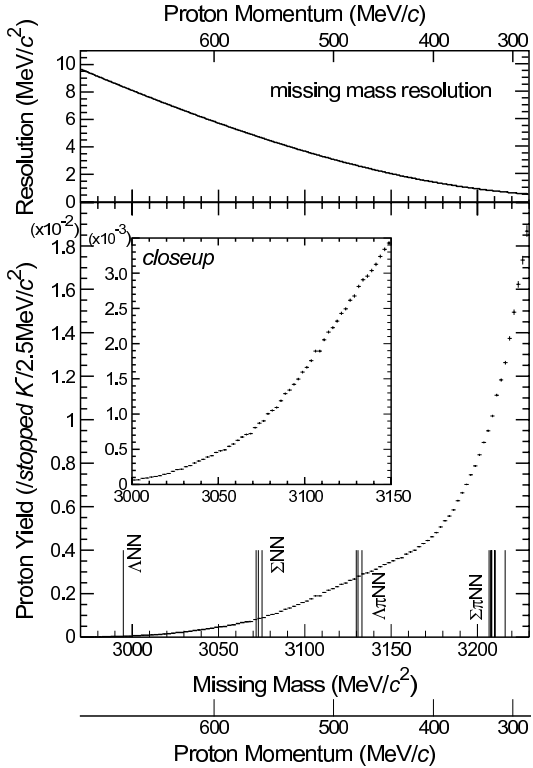}
\caption{The missing mass spectrum from the $^4$He$(K^{-}_{stopped},~p)$ reaction by the inclusive measurement. The abscissa is the missing mass and the ordinate is the proton yield normalized by the number of stopped kaons. The systematic error of the ordinate contains a 14 percent relative error. The upper figure shows the overall missing mass resolution in the present experiment.} 
\label{mass}
\end{center}
\end{figure}

Figure \ref{mass} shows the normalized missing mass spectrum obtained from the inclusive $^4$He$(K^{-}_{stopped},~p)$ reaction together with the proton momentum scale and energy thresholds of the $YNN$ and $Y\pi NN$. 
The ordinate unit shows the proton yield per stopped $K^-$, and the systematic error of the ordinate corresponds to 14 \% in the relative error,  which is dominated by the error of $f_{decay}$.  
An inset in the figure is the close-up view of the lower mass region.

The upper figure in Fig \ref{mass} shows the missing mass resolution as a function of the missing mass. It was derived from a Monte Carlo simulation with an assumption that the TOF resolution of protons is equivalent to that of $K^{-}_{\mu2}$ muons. This resolution is about twice better than that of Ref. \cite{E471p}. 
This was used in a spectral fitting for the determination of upper limits described in the following section.

\subsection{Upper limit of the formation branching ratio for a discrete state}
For quantifying the search results, 
we derived upper limits for the formation branching ratio of a tribaryon state. They were obtained by fitting the spectrum shown in Fig. \ref{mass} with an assumed peak  and a smooth background function.
 The fitting procedure is as follows;
\begin{itemize}
\item
 The fitting function was set to be a Voigtian (convolution of Gaussian and Lorentzian) as a peak and a polynomial function as a background. 
The yield of the state was determined by the area of the Lorentzian.
The natural width and center of the peak were fixed for each fitting.
The standard deviation of the Gaussian was set to be the experimental resolution taking the mass dependence into account as shown in Fig. \ref{mass}.
\item
 Since the background shape cannot be fully reproduced by a Monte Carlo simulation, we assume that the background can be expressed as a smooth function within the fit region.
\item
We tested three fit regions -- determined by the width of the peak function to be ($\pm n\times$ HWHM ($n~=~3,4,5$)).
Conservative upper limits were determined using the highest value of the results from each fit region.
\end{itemize}
We used this fit method with widths of 0, 20 and 40 MeV/$c^2$ by changing the mass from 3000 to 3200 MeV/$c^2$.

Systematic errors, mainly dominated by an uncertainty in $f_{decay}$,  were taken into account in the total error of the peak intensity  by summing 
quadratically the systematic error and fitting error
($\sigma_{total} = \sqrt{\sigma^2_{sys.} + \sigma^2_{stat.}}$).
In the fitting, the yield of the state sometimes results in an unphysical value, namely, a negative  Lorentzian area. 
In that case, we followed a prescription of {\it ``Unified Approach''} assuming Gaussian statistics \cite{fc}. 
The upper limit on the formation branching ratio at 95 \% confidence level 
was obtained using the table X in Ref. \cite{fc} with the resultant peak yield and the total error.  

\begin{figure}[hbt]
\begin{center}
 \includegraphics[width=0.5\columnwidth]{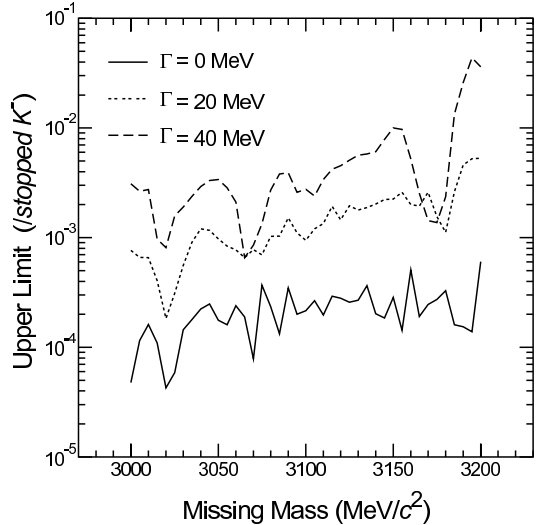}
\caption{Upper limits of the formation branching ratio for a strange tribaryon state at the 95 \% C.L. as a function of the missing mass.  Solid, dotted and dashed lines correspond to the assumed width of $\Gamma$ = 0, 20 and 40 MeV/$c^2$ states, respectively.} 
\label{ul}
\end{center}
\end{figure}

Figure \ref{ul} shows a plot of 
 the upper limits versus the missing mass for widths of the assumed states of 0, 20 and 40 MeV/$c^2$.
The present procedure gives a higher upper limit especially for the state with a larger width. 
Thus, it must be emphasized that the present procedure using  the inclusive proton spectrum only is insensitive for a broad peak structure, which will be investigated with exclusive measurements.

The present upper limit of the formation branching ratio at the mass of 3115 MeV/$c^2$, where the E471 group observed S$^0(3115)$, is $\sim$0.2 \%/($stopped~K^-$) with the width of $\Gamma$ = 20 MeV/$c^2$. Although the trigger condition was different, the E471 group reported a formation branching ratio of about 1 \%/($stopped~K^-$) in Ref. \cite{E471p}, which can be excluded by more than  a 95 \% confidence level.
We studied the reason for this discrepancy and found as a most likely cause for a fake peak formation, an erroneous slewing correction of the proton time of flight spectrum 
for the large proton signals compared to those of well calibrated minimum ionizing particles \cite{HYPmi}.

It is worth mentioning that the inclusive proton momentum spectrum from $^4$He($K^{-}_{stopped}$, $p$) has no structure whereas the FINUDA group found a peak in the proton momentum spectrum for $^6$Li($K^{-}_{stopped}$, $p$).
There is a controversial discussion concerning the production of monochromatic proton by antikaon absorption on low momentum deuteron clusters in nuclei
\cite{OT,AY07}. In the case of $^6$Li, the peak observed in the FINUDA spectrum may be attributed to such a process, however, it is obvious that the monochromatic process does not occur in the $K^- + ^4$He absorption case at the level of $1 \times 10^{-3}/(stopped~K^-)$.

\section{Conclusion}
In conclusion, we have searched for neutral tribaryon states with strangeness $-$1 and isospin 1 by missing-mass analysis of the inclusive $^4$He$(K^{-}_{stopped},~p)$ reaction. 
Quite high statistics for protons (more than one million) was accumulated in the momentum range of 300 -- 700 MeV/$c$. 
No significant narrow structure was observed. 
The upper limits for the formation branching ratio in the mass range
3000 $<$ M $<$ 3200 MeV/$c^2$ with an assumed width of 0, 20 and 40 MeV/$c^2$
 were determined to be approximately $2\times10^{-4}$, $1\times10^{-3}$ and $5\times10^{-3}$/($stopped~K^{-}$), respectively, with a 95 \% confidence level.
This result clearly indicates that a large formation ratio (order of 1 \%/($stopped~K^-$)) of a narrow tribaryon state, such as S$^0(3115)$ is excluded. 

We would like to thank all of the staff in KEK-PS for their continuous support. This research was partially supported by RIKEN, KEK and the Ministry of Education, Science, Sports and Culture, Grand-in-Aid for Scientific Research (S), 14102005.


\begin{thebibliography}{00}
\bibitem{AY02}    Y. Akaishi and T. Yamazaki, Phys. Rev. C 65 (2002) 044005.
\bibitem{E471p}   T. Suzuki et al., Phys. Lett. B 597 (2004) 26.
\bibitem{E471n}   M. Iwasaki et al., arXiv:nucl-ex/0310018.
\bibitem{ADY05}   Y. Akaishi, A. Dote and T. Yamazaki, Phys. Lett. B 613 (2005) 140.
\bibitem{Nona}    Y. Maezawa  et al., Prog. Theor. Phys. 114 (2005) 317.
\bibitem{OT}      E. Oset and H. Toki, Phys. Rev. C 74:015207, 2006, nucl-th/0509048.
\bibitem{Finuda}  M. Agnello et al. (FINUDA Collaboration), Nucl. Phys. A 775 (2006) 35.
\bibitem{AY07}    T. Yamazaki and Y. Akaishi, nucl-ex/0609041, Nucl. Phys. A 792 (2007) 229.
\bibitem{E471nim} M. Iwasaki et al., Nucl. Inst. Meth. A 473 (2001) 286.
\bibitem{meta}    T. Yamazaki et al., Phys. Rev. Lett. 63 (1989) 1590. 
\bibitem{OutaD}   H. Outa, Doctoral Dissertation, Univ. of Tokyo (2003).
\bibitem{fc}      G. J. Feldman and R. D. Cousins, Phys. Rev. D 57 (1998) 3873. 
\bibitem{HYPmi}   M. Iwasaki et al., arXiv:0706.0297 [nucl-ex], to be published in EPJ.  



\end{thebibliography}
\end{document}